\documentclass{zlab}

\usepackage[T1]{fontenc}
\usepackage{tgpagella}
\usepackage{mathpazo}
\usepackage{inconsolata}

\usepackage{xspace}
\usepackage{graphicx,amsmath,amssymb,hyperref}
\usepackage[authoryear,round]{natbib}
\usepackage{xcolor}

\usepackage{multirow}
\usepackage{longtable}
\usepackage{listings}
\usepackage{wrapfig}
\newcommand{\bench}{SWEeper-Bench}

\hypersetup{
  colorlinks=true,
  urlcolor=link,
  linkcolor=black,
  citecolor=link,
  filecolor=black,
  pdfborder={0 0 0}
}

\definecolor{link}{HTML}{0063BE}
\providecommand{\equationname}{Equation}
\providecommand{\sectionname}{Section}
\providecommand{\appendixrefname}{Appendix}

\newcommand{\figref}[1]{%
  \figurename~\hyperref[#1]{\textcolor{link}{\ref*{#1}}}%
}
\newcommand{\tabref}[1]{%
  \tablename~\hyperref[#1]{\textcolor{link}{\ref*{#1}}}%
}
\newcommand{\eqrefc}[1]{%
  \equationname~\hyperref[#1]{\textcolor{link}{\ref*{#1}}}%
}
\newcommand{\secrefc}[1]{%
  \sectionname~\hyperref[#1]{\textcolor{link}{\ref*{#1}}}%
}
\newcommand{\apprefc}[1]{%
  \appendixrefname~\hyperref[#1]{\textcolor{link}{\ref*{#1}}}%
}

\newcommand{\github}{\raisebox{-1.3pt}{\includegraphics[height=1.05em]{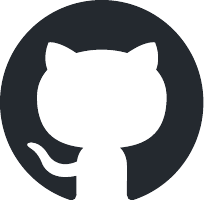}}\xspace}
\newcommand{\worldwideweb}{\raisebox{-1.3pt}{\includegraphics[height=1.05em]{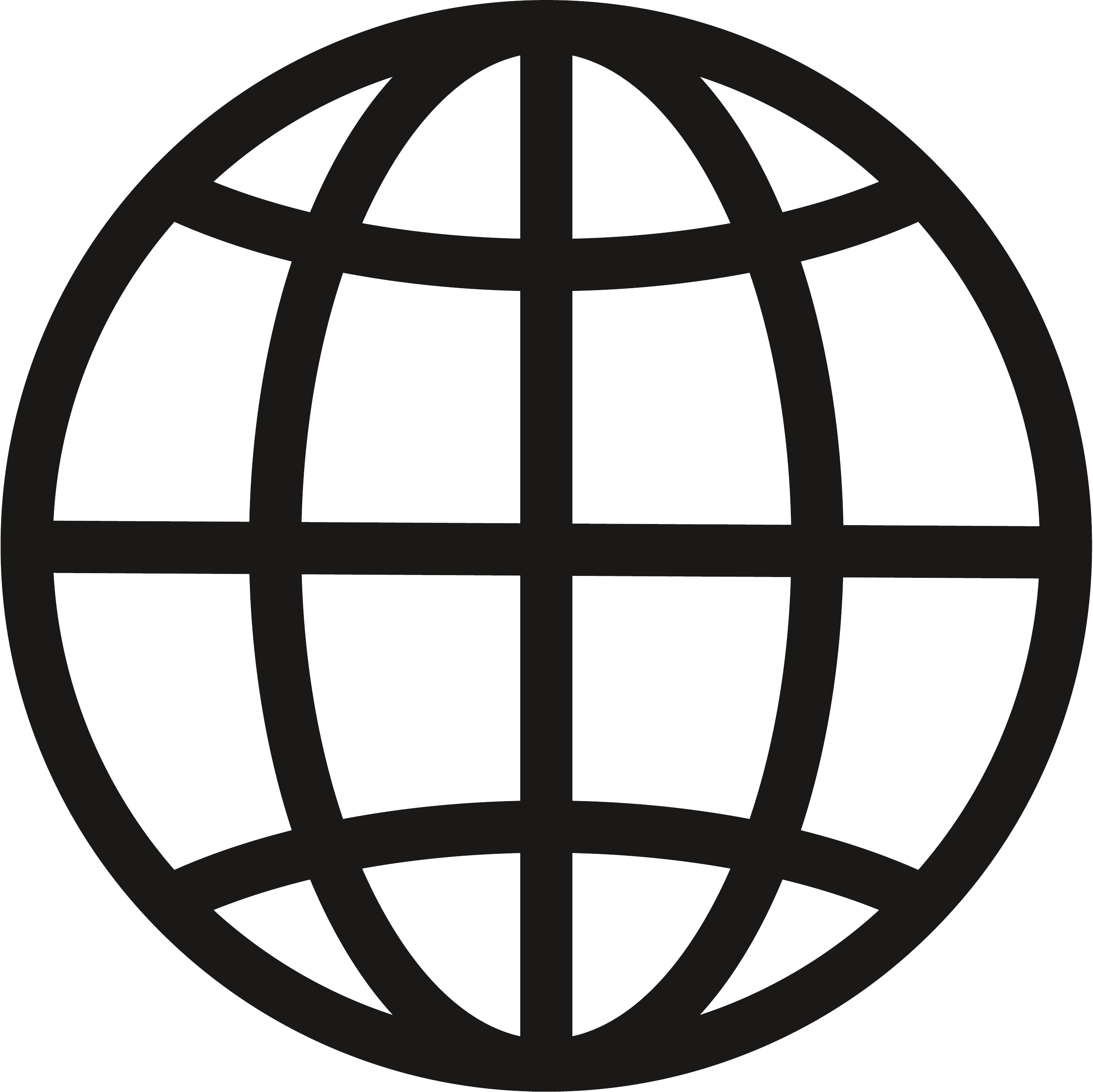}}\xspace}
\newcommand{\hf}{\raisebox{-1.3pt}{\includegraphics[height=1.05em]{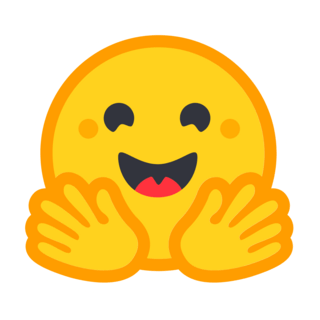}}\xspace}

\makeatletter
\fancypagestyle{firststyle}{
  \fancyhead[L]{}
  \fancyhead[C]{}
  \fancyhead[R]{}
  \fancyfoot[L]{}
  \fancyfoot[R]{}
}
\makeatother

\DeclareTextFontCommand{\textbf}{\bfseries}
\title{\centering \fontsize{18}{24}\selectfont{\texorpdfstring{\raisebox{-0.3\height}{\includegraphics[height=1.6em]{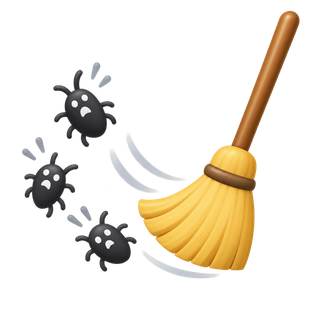}}\hspace{0.3em}}{}SWEeper-Bench: Can Agents Discover Bugs\\in Interactive Software?}}

\author{
    \vspace{.1cm}
    \parbox{\textwidth}{\centering
        Yang Yao\textsuperscript{1}$^{*}$ \quad
        Haozhe Chen\textsuperscript{1}$^{*}$ \quad
        Bingyi Kang \quad
        Karthik R Narasimhan\textsuperscript{1} \quad
        Zhuang Liu\textsuperscript{1}
    }
    \\
    \vspace{.15cm}
    {\normalfont\fontsize{11}{15}\selectfont {\textsuperscript{1}Princeton University}}
}

\makeatletter
\renewcommand{\abscontent}{}%
\makeatother

\newenvironment{abstractblock}{%
  {\centering\large\bfseries Abstract\par}
  \vspace{0.2em}
  \begin{list}{}{%
      \setlength{\leftmargin}{2em}
      \setlength{\rightmargin}{2em}
      \setlength{\topsep}{0pt}
      \setlength{\parsep}{0pt}
  }
  \item[]
}{%
  \end{list}
  \par\normalfont\vspace{1em}
}

\begin{document}

\begingroup
\makeatletter
\let\raggedright\centering
\makeatother

\maketitle
\endgroup

\setlength{\skip\footins}{0.3cm}
\begingroup\renewcommand\thefootnote{}\footnotetext{\fontsize{9}{11}\selectfont$^*$\,Equal Contribution}\addtocounter{footnote}{-1}\endgroup

\vspace{-1cm}
\begin{center}
    \worldwideweb \href{https://sweeperbench.com}{\textbf{Website}} \quad
    \github \href{https://github.com/zlab-princeton/sweeper-bench}{\textbf{Code}} \quad
    \hf \href{https://huggingface.co/datasets/zlab-princeton/SWEeper-Bench}{\textbf{Data}} \quad
    \hf \href{https://huggingface.co/datasets/zlab-princeton/SWEeper-Bench-traces}{\textbf{Agent Traces}}
\end{center}

\newcommand{\abstractcontent}{%
Coding agents increasingly interact with users by generating interactive software on the fly rather than replying in text, yet many bugs in such software reveal themselves only after multi-step interactions. Today, users must discover these bugs and wait for fixes, making the development loop slow and expensive. Existing benchmarks typically assume the bugs have already been identified, leaving agents' ability to find interaction-dependent failures largely unevaluated. We introduce \bench{}, 200 tasks built from real bug fixes in 200 widely used open-source web applications. Each task gives the agent only the codebase and an open-ended request such as ``Find and fix issues in Focalboard's board sharing and membership.'' Because these bugs are hard to detect from code alone, agents must design interactive tests to uncover them. An agentic verifier then judges each repair by what a user would observe in the browser. The best of 15 frontier agents passes only 59.0\% of tasks. Through stage-by-stage analysis of agent trajectories, we show that the biggest bottleneck is identifying the buggy behavior: agents often reach the relevant feature but never trigger the failure or fail to recognize it. \bench{} measures progress toward agents that close this loop themselves, delivering software that works without waiting for user feedback.
}

\vspace{0cm}

\begin{figure}[!ht]
    \centering
    \includegraphics[width=\textwidth]{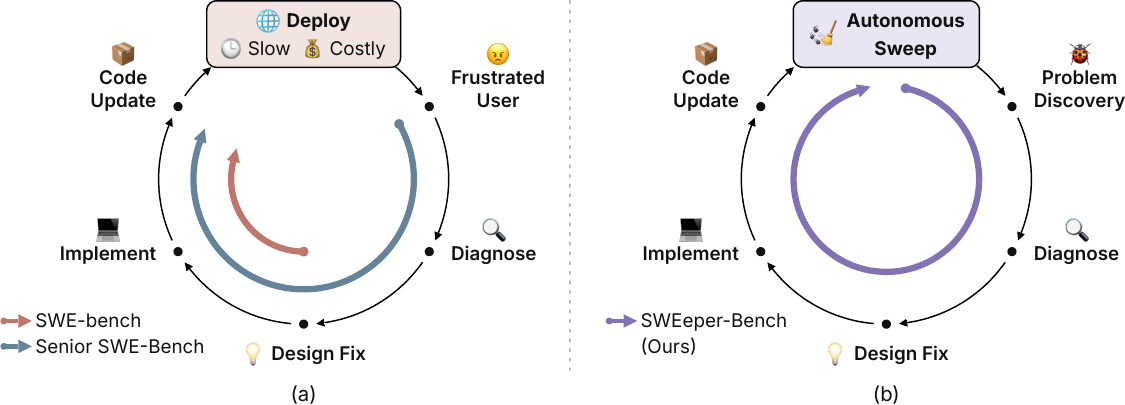}
    \caption{\textbf{From waiting for bug reports to sweeping for bugs.} (a) Existing benchmarks start after a user reports a bug, which comes only after a slow, costly deployment. (b) In \bench{}, the agent must discover the problem itself by exploring the code and the running application, then diagnose and fix it before users encounter it.}
    \label{fig:motivation}
\end{figure}

\vspace{-0.1cm}

\begin{abstractblock}
\abstractcontent
\end{abstractblock}

\newpage

\section{Introduction}

\begin{figure}[!t]
    \centering
    \includegraphics[width=\textwidth]{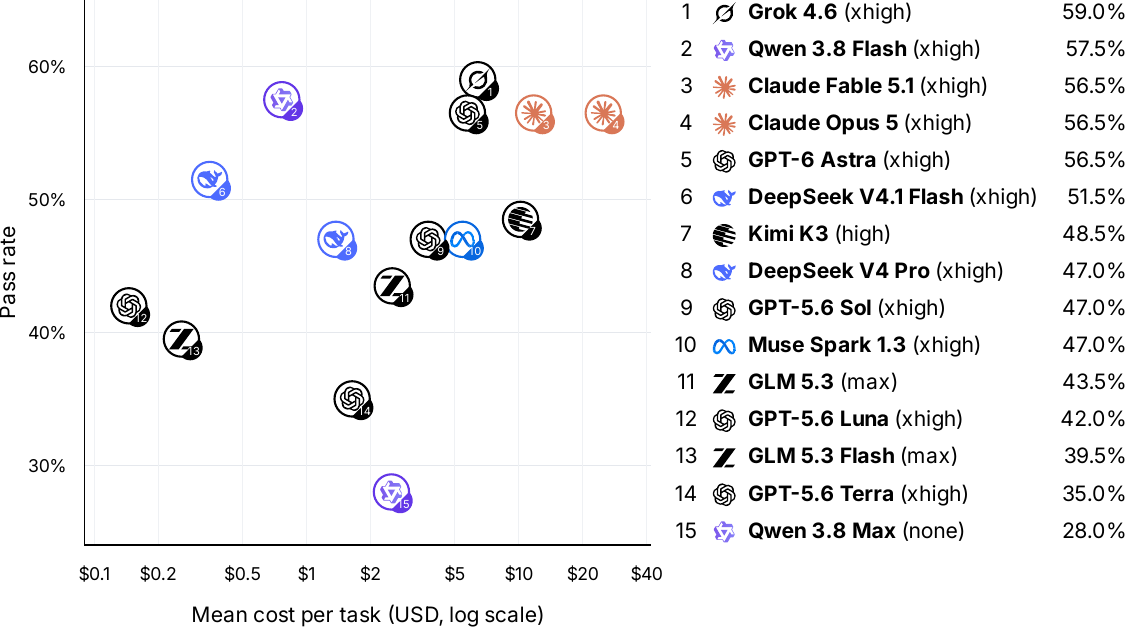}
    \caption{\textbf{\bench{} is far from solved.} Pass rate versus mean prediction cost per task for 15 agents. Each agent pairs a model with a harness. A task passes only if both its target and preservation behavior tests pass. Nine agents pass fewer than half of the tasks, and agents with similar pass rates differ in cost by more than 30 times. \tabref{tab:main} gives the full results.}
    \label{fig:leaderboard}
\end{figure}

Coding agents such as Codex~\citep{openai2026codex}, Claude Code~\citep{anthropic2026claudecode}, and Cursor~\citep{cursor2026} can now build interactive software from short requests, but such software is hard to debug. Its bugs often appear only after a specific sequence of user interactions: a board shared with a colleague never shows up in the colleague's sidebar, or a coupon accepted in the cart is rejected at checkout. Because these failures depend on how people use the software, they are hard to spot from code alone and usually surface only when people use it in the wild. Repair therefore stays reactive: users run into problems and report them, and agents must wait for this feedback before they can improve the software (\figref{fig:motivation}a).

This loop matters more as coding agents increasingly generate interactive software on the fly instead of replying in text. Imagine asking an assistant to organize a team offsite. Instead of replying in text, it builds a small app with a sign-up form, a shared schedule, and a budget tracker, and shares it with twenty colleagues within minutes. No testing team ever sees this app, so if some sign-ups silently vanish, the colleagues discover the bug themselves and wait for a fix. Routing every bug through users is slow and expensive, and users become the bottleneck of the whole loop. Agents must instead close the loop on their own and sweep out as many bugs as possible before presenting their software. We therefore ask: \emph{can agents discover and fix the problems users will encounter in interactive software on their own?}

Existing benchmarks typically assume the bug has already been identified, leaving agents' ability to find interaction-dependent failures largely unevaluated. SWE-bench~\citep{jimenez2024swe} and its successors~\citep{yang2024swe,rashid2025swe,deng2025swe,zan2026multi,zhang2026swe} start from an issue that a user has already reported, so they measure diagnosis and repair but not discovery. They also grade repairs with hidden unit tests~\citep{jimenez2024swe}, which check internal state rather than what a user sees and can reject valid fixes that take a different approach.

We introduce \bench{}, 200 tasks that test whether agents can sweep the bugs out of real web applications (\figref{fig:motivation}b). Each task comes from a real bug fix in a different open-source application, and its bug appears only after a sequence of user interactions. The agent receives the codebase, the running application with a browser, and an open-ended request that names only a product area. To succeed, it must understand how the application is used, anticipate what could go wrong, and design interactive tests that expose the bug before fixing it. An agentic verifier then operates the repaired application in the browser and judges only what a user would observe, so any correct implementation passes.

Across 15 frontier agents, the best passes only 59.0\% of tasks, and nine pass fewer than half (\figref{fig:leaderboard}). Agents can fix these bugs once someone finds them: when we replace the open-ended request with a detailed bug description, all five agents we test pass about 96\% of tasks. Through stage-by-stage analysis of agent trajectories, we show that the biggest bottleneck is identifying the buggy behavior. Agents reach the feature containing the bug in 75.4\% of trajectories but identify the buggy behavior in only 50.0\%: they either never trigger the failure or do not recognize it. They also rarely test through the browser unless told to.

\bench{} moves the evaluation of coding agents from repairing bugs to discovering them, a step toward agents that close the loop of building, testing, and fixing software on their own.

\section{Constructing \bench{}}
\label{sec:construction}

\begin{figure}[!t]
    \centering
    \includegraphics[width=\textwidth]{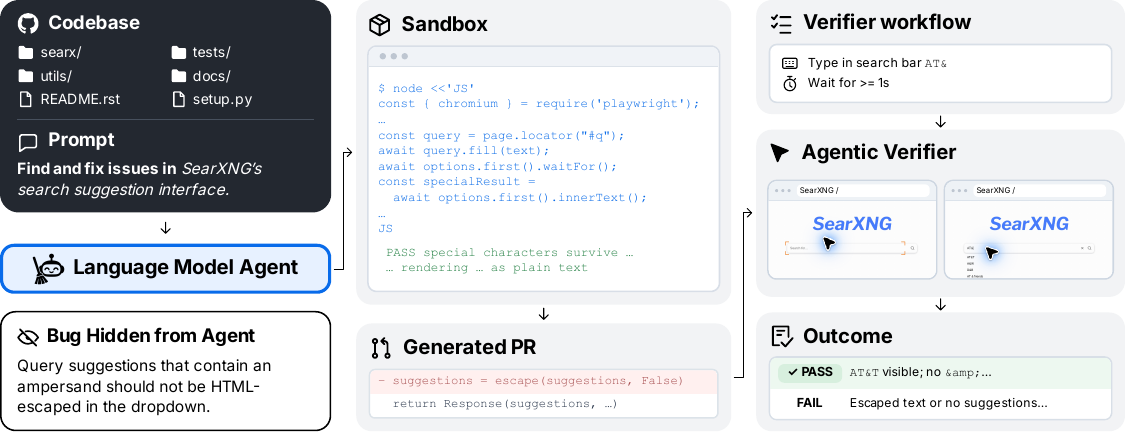}
    \caption{\textbf{Task execution and verification.} Given only the buggy codebase and an open-ended prompt, the agent explores the code and the running application in a sandbox, discovers the bug, and produces a patch. An agentic verifier applies the patch to a fresh copy of the application and runs predefined behavior tests through the browser. It judges only what a user would observe.}
    \label{fig:pipeline}
\end{figure}

Each \bench{} task pairs a buggy application and an open-ended prompt with two behavior tests that an agentic verifier runs in the browser (\figref{fig:pipeline}). We describe how we curate tasks (\secrefc{sec:curation}), what agents receive (\secrefc{sec:prompt}), and how we verify repairs (\secrefc{sec:verifier}).

\subsection{Curating Tasks}
\label{sec:curation}

\begin{wrapfigure}{r}{0.5\textwidth}
    \vspace{2pt}
    \centering
    \includegraphics[width=\linewidth]{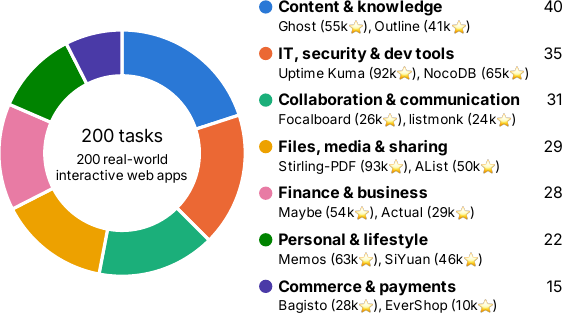}
    \caption{\textbf{200 tasks from 200 real-world web applications.} Each task comes from a different open-source application, grouped here into seven domains. Examples are the two most-starred repositories in each domain.}
    \label{fig:distribution}
    \vspace{-10pt}
\end{wrapfigure}

We build each task from a merged pull request that fixes a user-facing bug in an open-source web application: the parent commit is the buggy version, and the merged fix is the reference repair. Each application contributes one task. The 200 tasks thus span 200 applications in seven domains (\figref{fig:distribution}). The median repository has 4.6k GitHub stars.

We select bugs that are hard to find but obvious once seen: they are reproducible through normal use, and any user would agree the behavior needs fixing. An agent may fix several issues in the named area, but we evaluate only the target bug. A complete sweep of the area should include it.

\figref{fig:focalboard-procedure} shows a typical task. Focalboard is a project board tool. When a user shares a board, it never appears in the recipient's sidebar because adding a member saves the membership but never files the board under the member's default sidebar category. Nothing looks wrong to the user who shares the board, and the recipient can still open it through a direct link. The failure appears only after nine steps across two accounts. The fix is a single function call.

We reproduce every bug by hand in the browser, confirm that the reference fix resolves it, and walk through both behavior tests. The agent never sees the original issue, pull request, reference patch, repository history, or behavior tests, and outbound network access is blocked.

\subsection{Open-Ended Prompt}
\label{sec:prompt}

Each task gives the agent a one-sentence request that names a product area, such as ``Find and fix issues in Focalboard's board sharing and membership.'' It does not say what is broken, where the code lives, or how to reproduce the problem. This mirrors asking a colleague to sweep a feature before release. The agent works in an isolated sandbox with the buggy source code, a running instance that reflects its edits, a command interface for tests and builds, and a Chromium browser with Playwright. It decides for itself what to read, which scenarios to test, and when to stop. Our default prompt also asks the agent to test the application in the browser like a real user (\apprefc{app:prompts}), and \secrefc{sec:browser-prompt} shows why this matters. When the agent stops, we evaluate only its code changes.

\begin{table}[!t]
\centering
\small
\begingroup
\newcommand{\sweepermodelicon}[1]{\makebox[1.35em][l]{\raisebox{-0.18ex}{\includegraphics[width=1em,height=1em,keepaspectratio]{icons/#1.pdf}}}}
\newcommand{\sweepermodelname}[2]{\sweepermodelicon{#1}#2}
\setlength{\tabcolsep}{4pt}
\begin{tabular}{lccccc}
\toprule
& \multicolumn{2}{c}{Buggy version} & \multicolumn{2}{c}{Reference fix} & \\
\cmidrule(lr){2-3}\cmidrule(lr){4-5}
Verifier & Target $\downarrow$ & Preservation $\uparrow$ & Target $\uparrow$ & Preservation $\uparrow$ & Avg. cost \\
\midrule
\sweepermodelname{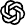}{GPT-5.6 Luna, Codex} & 0.2 $\pm$ 0.3 & 99.2 $\pm$ 0.8 & 99.8 $\pm$ 0.3 & 99.3 $\pm$ 0.3 & \$0.160 \\
\sweepermodelname{openai}{GPT-5.6 Luna, Browser Use} & 0.8 $\pm$ 0.3 & 99.8 $\pm$ 0.3 & 99.8 $\pm$ 0.3 & 99.7 $\pm$ 0.3 & \$0.635 \\
\sweepermodelname{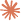}{Claude Sonnet 5, Browser Use} & 0.2 $\pm$ 0.3 & 99.8 $\pm$ 0.3 & 99.8 $\pm$ 0.3 & 100.0 $\pm$ 0.0 & \$3.840 \\
\bottomrule
\end{tabular}
\caption{\textbf{The verifier is accurate and consistent.} Pass rate (\%) of each behavior test on the buggy version and the reference fix, mean $\pm$ standard deviation over three runs of 200 tasks. Arrows mark the ideal: 0\% for the target test on the buggy version, 100\% otherwise. Avg. cost is USD per task. Verifiers use GPT-5.6 Luna~\citep{openai2026gpt56} or Claude Sonnet 5~\citep{anthropic2026sonnet5} with Codex~\citep{openai2026codex} or Browser Use~\citep{browser_use2024}.}
\label{tab:robustness}
\endgroup
\end{table}

\subsection{Agentic Verifier}
\label{sec:verifier}

\begin{figure}[!t]
    \centering
    \includegraphics[width=\textwidth]{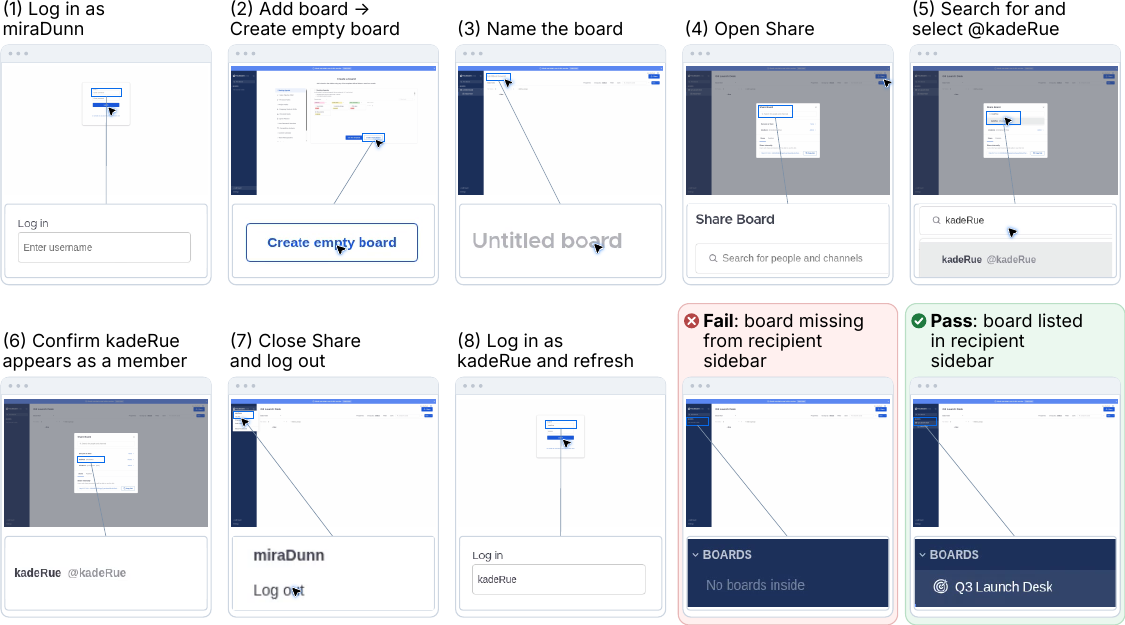}
    \caption{\textbf{A bug that takes nine steps and two accounts to see.} In Focalboard, the owner creates and names a board, shares it with a second user, and logs out (1 to 7). The second user then logs in and refreshes (8). With the bug, the shared board is missing from the recipient's sidebar (Fail); with the fix, it appears (Pass). The root cause is that adding a member saves the membership but never adds the board to the member's default sidebar category.}
    \label{fig:focalboard-procedure}
\end{figure}

We verify repairs the way a user would notice them. We apply the agent's code changes to a fresh copy of the buggy application, and a verifier agent then uses it through the browser without seeing the source code. Each task has two behavior tests, written as step-by-step user instructions with explicit pass and fail criteria (\apprefc{app:workflow}). The target behavior test reproduces the interactions that expose the bug and checks that the intended behavior is restored; the preservation behavior test checks that nearby features still work. A task passes only if both tests pass.

To measure reliability, we run both tests on the buggy version and on the reference fix with three verifier configurations, three times each. An accurate verifier passes the target test on none of the buggy versions and all of the reference fixes, and passes the preservation test on both. All configurations come within one percentage point of these ideal rates (\tabref{tab:robustness}). Across runs, the standard deviation is at most 0.8 points. We use the cheapest configuration, GPT-5.6 Luna with Codex, as the default verifier.

\begin{table}[!t]
\centering
\small
\begingroup
\newcommand{\sweepermodelicon}[1]{\makebox[1.35em][l]{\raisebox{-0.18ex}{\includegraphics[width=1em,height=1em,keepaspectratio]{icons/#1.pdf}}}}
\newcommand{\sweepermodelname}[2]{\sweepermodelicon{#1}#2}
\setlength{\tabcolsep}{3.5pt}
\begin{tabular}{lllccccr}
\toprule
Model & Harness & Reasoning & Pass $\uparrow$ & Target $\uparrow$ & Pres. $\uparrow$ & Time (min) & Cost (\$) \\
\midrule
\sweepermodelname{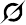}{Grok 4.6} & Cursor & xhigh & \textbf{59.0} & 59.0 & 100.0 & 12.3 $\pm$ 3.9 & 6.41 \\
\sweepermodelname{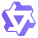}{Qwen 3.8 Flash} & Codex & xhigh & 57.5 & 58.0 & 97.5 & 101.8 $\pm$ 46.6 & 0.76 \\
\sweepermodelname{claude}{Claude Fable 5.1} & Claude Code & xhigh & 56.5 & 56.5 & 100.0 & 28.2 $\pm$ 15.8 & 11.76 \\
\sweepermodelname{claude}{Claude Opus 5} & Claude Code & xhigh & 56.5 & 57.5 & 98.5 & 42.8 $\pm$ 23.2 & 25.01 \\
\sweepermodelname{openai}{GPT-6 Astra} & Codex & xhigh & 56.5 & 56.5 & 100.0 & 13.7 $\pm$ 3.4 & 5.73 \\
\sweepermodelname{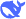}{DeepSeek V4.1 Flash} & DeepSeek & xhigh & 51.5 & 51.5 & 99.0 & 41.8 $\pm$ 20.0 & 0.35 \\
\sweepermodelname{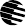}{Kimi K3} & Kimi Code & high & 48.5 & 49.0 & 99.5 & 90.1 $\pm$ 63.1 & 10.19 \\
\sweepermodelname{deepseek}{DeepSeek V4 Pro} & DeepSeek & xhigh & 47.0 & 47.0 & 99.5 & 58.2 $\pm$ 25.8 & 1.38 \\
\sweepermodelname{openai}{GPT-5.6 Sol} & Codex & xhigh & 47.0 & 47.0 & 98.0 & 16.8 $\pm$ 5.8 & 4.28 \\
\sweepermodelname{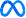}{Muse Spark 1.3} & Muse Code & xhigh & 47.0 & 47.5 & 99.5 & 25.6 $\pm$ 20.3 & 4.75 \\
\sweepermodelname{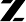}{GLM 5.3} & Codex & max & 43.5 & 43.5 & 99.0 & 20.3 $\pm$ 10.9 & 2.53 \\
\sweepermodelname{openai}{GPT-5.6 Luna} & Codex & xhigh & 42.0 & 43.5 & 98.5 & 16.5 $\pm$ 6.0 & 0.15 \\
\sweepermodelname{zai}{GLM 5.3 Flash} & Codex & max & 39.5 & 39.5 & 99.0 & 22.9 $\pm$ 11.7 & 0.26 \\
\sweepermodelname{openai}{GPT-5.6 Terra} & Codex & xhigh & 35.0 & 35.5 & 99.0 & 22.7 $\pm$ 7.0 & 1.64 \\
\sweepermodelname{qwen}{Qwen 3.8 Max} & Codex & none & 28.0 & 28.0 & 99.0 & 24.7 $\pm$ 13.5 & 2.51 \\
\bottomrule
\end{tabular}
\caption{\textbf{Results on \bench{}.} Pass rates are percentages of the 200 tasks. Pass requires both behavior tests to pass; Target and Pres. (preservation) report each test separately. Time is mean $\pm$ standard deviation of agent work time per task; cost is mean prediction cost per task.}
\label{tab:main}
\endgroup
\end{table}

\begin{figure}[!t]
    \centering
    \includegraphics[width=\textwidth]{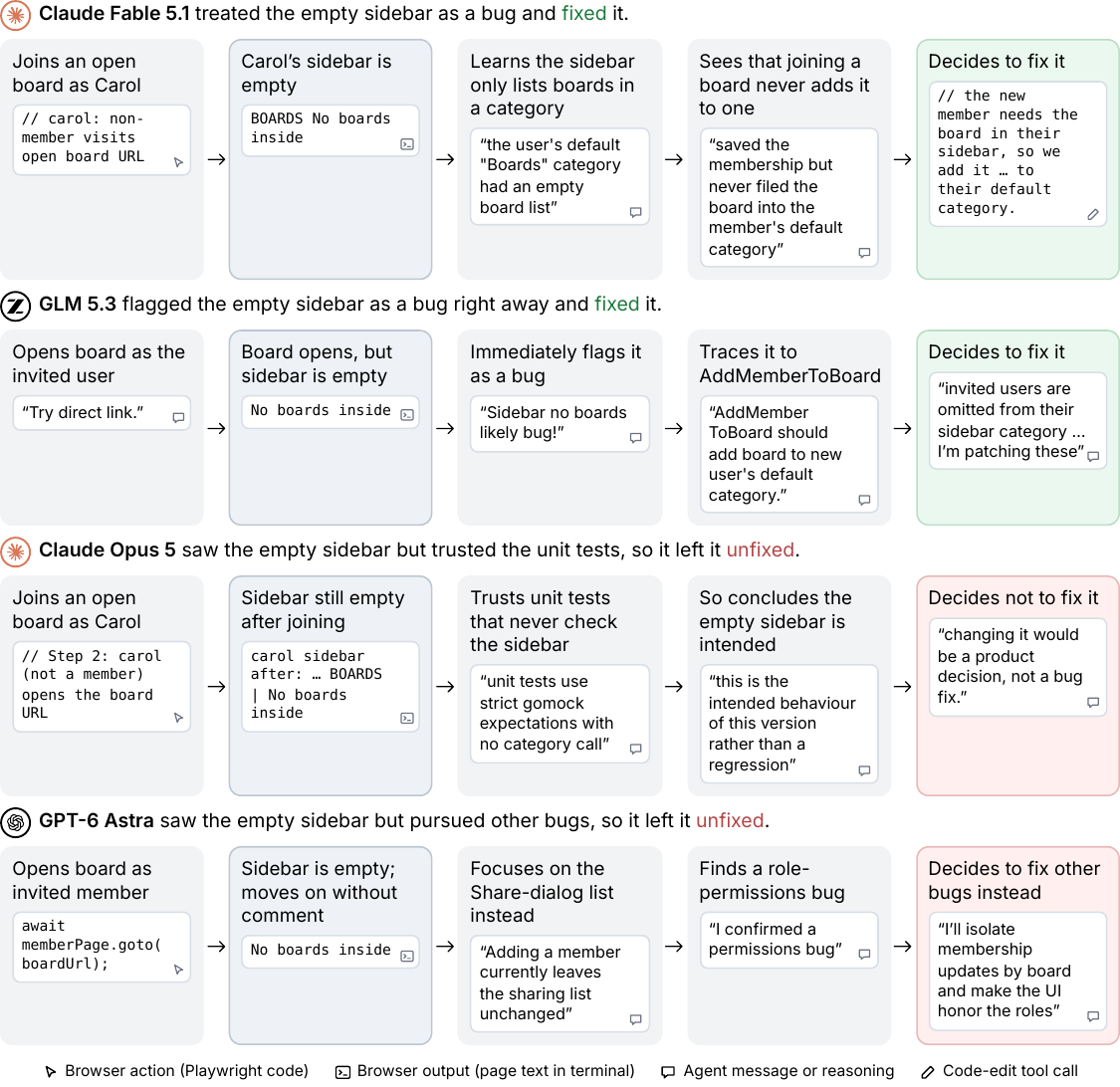}
    \caption{\textbf{Seeing a bug is not the same as recognizing it.} Four agents on the Focalboard task all reach the empty recipient sidebar (shaded column) but respond differently. Claude Fable 5.1 and GLM 5.3 treat it as a bug and fix it; Claude Opus 5 trusts the existing unit tests and judges it intended; GPT-6 Astra pursues other bugs.}
    \label{fig:focalboard-trajectories}
\end{figure}

\section{How Do Agents Perform?}
\label{sec:results}

\subsection{Benchmark Results}
\label{sec:main-results}

We evaluate 15 agents (\tabref{tab:main}). Each pairs a frontier model with its native harness if one exists and with Codex otherwise. We list the models~\citep{xai2026grok46,qwen2026qwen38,anthropic2026fable51,anthropic2026opus5,openai2026gpt6astra,openai2026gpt56,deepseek2026v41,deepseek2026v4,moonshot2026kimik3,meta2026musespark,zai2026glm53} and harnesses~\citep{openai2026codex,anthropic2026claudecode,cursor2026,moonshot2026kimicode,meta2026musecode,deepseek2026harness}. All runs use the browser instruction and an 8-hour limit per task.

\bench{} is far from solved. Grok 4.6 with Cursor passes 59.0\% of tasks, followed by Qwen 3.8 Flash at 57.5\% and three agents at 56.5\%; nine of fifteen pass fewer than half. Nearly all failures come from the target behavior test, while preservation pass rates stay between 97.5\% and 100\%. Agents rarely break working features but often miss the bug they were asked to find. Similar pass rates also hide very different time and cost: Grok 4.6 averages 12 minutes per task and Qwen 3.8 Flash 102 minutes, and Claude Opus 5 matches Claude Fable 5.1 at more than twice the cost.

\subsection{Where Do Agents Stumble?}
\label{sec:stages}

\begin{figure}[!t]
\centering
\begin{minipage}[b]{0.48\textwidth}\centering\includegraphics[width=\linewidth]{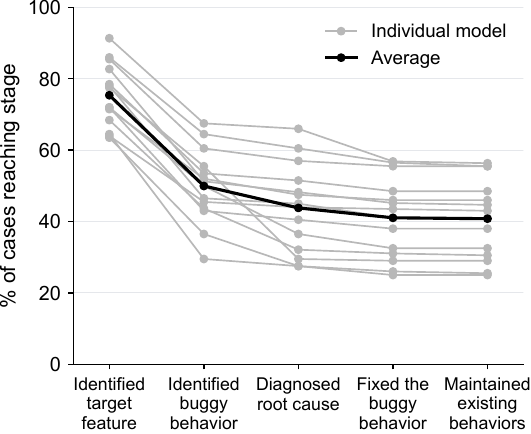}\end{minipage}\hfill
\begin{minipage}[b]{0.48\textwidth}\centering\includegraphics[width=\linewidth]{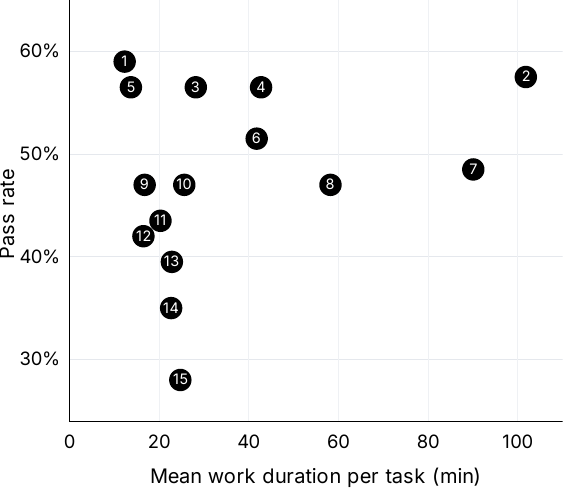}\end{minipage}
\par\vspace{2pt}
\begin{minipage}[t]{0.48\textwidth}
\caption{\textbf{The largest drop is at identifying the buggy behavior.} Share of trajectories reaching each stage. Each stage requires all previous ones.}
\label{fig:stages}
\end{minipage}\hfill
\begin{minipage}[t]{0.48\textwidth}
\caption{\textbf{Working longer does not lead to higher pass rates.} Pass rate versus mean agent work time per task. Numbers are agent ranks from \figref{fig:leaderboard} and \tabref{tab:main}.}
\label{fig:duration}
\end{minipage}
\end{figure}

Failing agents often get close to the bug (\figref{fig:focalboard-trajectories}). To locate where they stop, we split each trajectory into five cumulative stages (\figref{fig:stages}). GPT-5.6 Luna reads each trajectory and, using a task-specific description written from the reference fix, judges whether the agent identifies the target feature, the buggy behavior, and the root cause. The last two stages come from the behavior tests. Averaged over 14 agents, the share of trajectories falls from 75.4\% at the first stage to 40.8\% at the last.

\textbf{Finding the target feature.} Agents reach the feature that contains the bug in 75.4\% of trajectories (63.5\% to 91.4\% across agents). In the AbanteCart task, Grok 4.6 fixes phone numbers, addresses, and notifications on the account profile page but never tries the ID photo upload where the bug lives.

\textbf{Identifying the buggy behavior.} This is the largest drop, to 50.0\%. Agents often reach the feature but never run the interaction that exposes the bug: in the Answer task, Grok 4.6 accepts and reverts answers but never opens a question through its short-ID link. Even agents that trigger the failure may not recognize it (\figref{fig:focalboard-trajectories}). On the Focalboard task, all four agents we inspect reach the empty recipient sidebar. GPT-6 Astra moves on without comment, and Claude Opus 5 concludes the behavior is intended because the existing unit tests expect no sidebar update. Existing tests can thus reinforce the wrong expectation.

\textbf{Diagnosing the root cause.} This stage keeps 43.8\% of trajectories. In the SearXNG task, GLM 5.3 Flash sees ampersands escaped in search suggestions but blames client-side query encoding instead of server-side HTML escaping. In the Audiobookshelf task, Qwen 3.8 Flash reproduces the failed author merge but attributes it to case-sensitive lookup rather than duplicate insertion.

\textbf{Fixing and maintaining behavior.} Once the cause is found, agents usually fix it: 41.0\% of trajectories pass the target test, and 40.8\% also keep nearby features working. The remaining failures are rare regressions and incomplete fixes. For example, GLM 5.3 Flash corrects the EverShop variant editor's default stock but omits the API field that supplies the quantity.

\subsection{Behavioral Tendencies}
\label{sec:behavior}

\textbf{Longer work does not mean better results.} Agents differ in work time by nearly an order of magnitude, and this time does not track pass rate (\figref{fig:duration}). Grok 4.6 and GPT-6 Astra average about 13 minutes per task. Qwen 3.8 Flash and Kimi K3 average 102 and 90 minutes for similar or lower pass rates. Long runs can fail to converge: on three tasks, Qwen 3.8 Flash works for nearly the full 8 hours without producing a patch. Short runs risk missing the bug: Grok 4.6 stops after 7.8 minutes on the AbanteCart task.

\begin{figure}[!t]
    \centering
    \includegraphics[width=\textwidth]{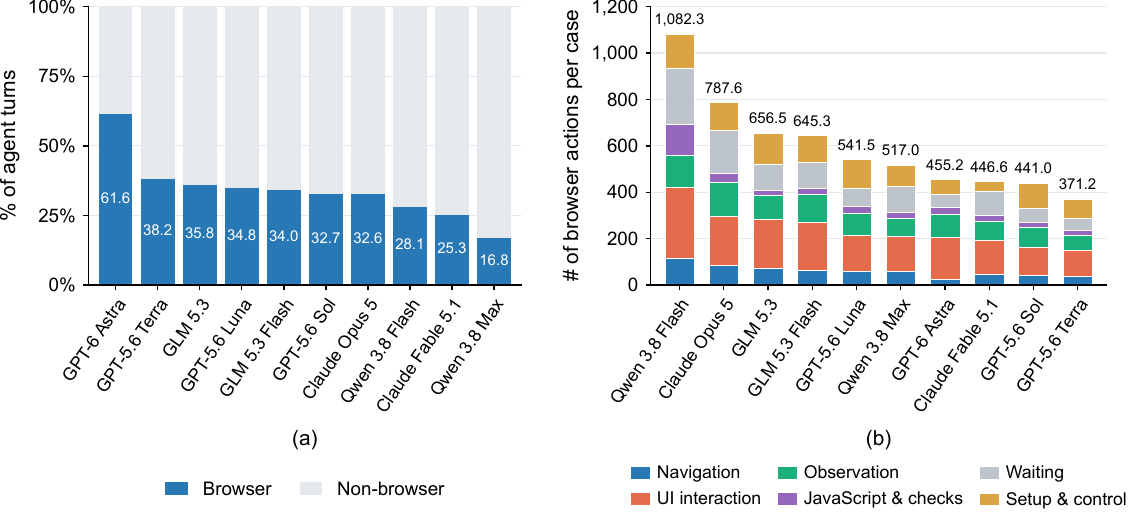}
    \caption{\textbf{How agents use the browser.} (a) Percentage of agent turns that use the browser. (b) Mean browser actions per task, grouped by type.}
    \label{fig:browser}
\end{figure}

\textbf{More browser use does not mean better results either.} Among agents whose harness records tool calls, browser actions make up 17\% to 62\% of turns, or 371 to 1,082 actions per task (\figref{fig:browser}). Neither tracks pass rate. Qwen 3.8 Max issues 517 browser actions per task and passes 28.0\%. Claude Fable 5.1 uses the browser in only 25\% of its turns and passes 56.5\%. What matters is which scenarios an agent tests and whether it recognizes a failure when it appears (\secrefc{sec:stages}).

\section{What Affects Performance?}
\label{sec:factors}

\subsection{Does Browser Prompting Help?}
\label{sec:browser-prompt}

\begin{wraptable}{r}{0.46\textwidth}
\vspace{-8pt}
\centering
\small
\begingroup
\newcommand{\sweepermodelicon}[1]{\makebox[1.35em][l]{\raisebox{-0.18ex}{\includegraphics[width=1em,height=1em,keepaspectratio]{icons/#1.pdf}}}}
\newcommand{\sweepermodelname}[2]{\sweepermodelicon{#1}#2}
\setlength{\tabcolsep}{4pt}
\begin{tabular}{llcc}
\toprule
Model & Instruction & Browser & Pass \\
\midrule
\multirow{2}{*}{\sweepermodelname{openai}{GPT-5.6 Luna}} & No & 2/200 & 39.5 \\
& Yes & 200/200 & \textbf{42.0} \\
\midrule
\multirow{2}{*}{\sweepermodelname{zai}{GLM 5.3}} & No & 6/200 & 31.5 \\
& Yes & 200/200 & \textbf{43.5} \\
\bottomrule
\end{tabular}
\captionof{table}{\textbf{Agents rarely test in the browser unless asked.} Browser: tasks where the agent drives a real browser.}
\label{tab:browser-prompt}
\endgroup
\vspace{-10pt}
\end{wraptable}
Our default prompt asks agents to test in the browser like a real user (\apprefc{app:prompts} gives the full template and the lines we remove). To see whether they would do so on their own, we remove this instruction for GPT-5.6 Luna and GLM 5.3 and keep everything else unchanged. The browser remains installed. The agents then almost never open the browser: GPT-5.6 Luna drives one on 2 of 200 tasks and GLM 5.3 on 6, compared with all 200 when instructed (\tabref{tab:browser-prompt}). Instead, they read code, run shell commands, and send HTTP requests, and pass rates drop from 42.0\% to 39.5\% and from 43.5\% to 31.5\%. Interactive testing is thus a capability agents do not yet reach for on their own. We keep the instruction so the benchmark measures how well agents test rather than whether they think to.

\subsection{Is Finding Harder Than Fixing?}
\label{sec:detailed-prompt}

\begin{wraptable}{r}{0.46\textwidth}
\vspace{-8pt}
\centering
\small
\begingroup
\newcommand{\sweepermodelicon}[1]{\makebox[1.35em][l]{\raisebox{-0.18ex}{\includegraphics[width=1em,height=1em,keepaspectratio]{icons/#1.pdf}}}}
\newcommand{\sweepermodelname}[2]{\sweepermodelicon{#1}#2}
\setlength{\tabcolsep}{4pt}
\begin{tabular}{lcc}
\toprule
Model & Open-ended & Detailed \\
\midrule
\sweepermodelname{deepseek}{DeepSeek V4.1 Flash} & 51.5 & \textbf{96.5} \\
\sweepermodelname{openai}{GPT-5.6 Sol} & 47.0 & \textbf{96.5} \\
\sweepermodelname{zai}{GLM 5.3} & 43.5 & 96.0 \\
\sweepermodelname{openai}{GPT-5.6 Luna} & 42.0 & 96.0 \\
\sweepermodelname{openai}{GPT-5.6 Terra} & 35.0 & 96.0 \\
\bottomrule
\end{tabular}
\captionof{table}{\textbf{Agents fix bugs reliably once told what the bug is.} Pass rate (\%) by prompt. GPT-5.6 Sol has 199 detailed-prompt tasks.}
\label{tab:detailed-prompt}
\endgroup
\vspace{-10pt}
\end{wraptable}
To separate finding a bug from fixing it, we give five agents from the main evaluation a detailed prompt that describes the target bug and its expected behavior (\apprefc{app:prompts}). This prompt plays the role of a bug report. All five then pass between 96.0\% and 96.5\% of tasks instead of 35.0\% to 51.5\%, and the gap between agents almost disappears (\tabref{tab:detailed-prompt}). Trajectories also narrow: for GPT-5.6 Luna, the median number of browser launches falls from 26 to 14, and the median number of changed lines from 49 to 13. The difficulty of \bench{} lies in producing the bug report, not in acting on it.

\subsection{Does the Harness Matter?}
\label{sec:harness}

Each agent in \tabref{tab:main} pairs a model with a harness. To isolate the effects of harnesses, we run GPT-5.6 Luna under five harnesses with all else held constant (\tabref{tab:harness}). Pass rates range from 35.0\% to 42.5\%, and Luna does best in Codex, its native harness. A well-matched harness helps the model use its capabilities, but the model itself accounts for most of the variation across agents.

\subsection{Does Working Longer Help?}
\label{sec:time-budget}

\begin{figure}[!t]
\newsavebox{\harnesstabbox}
\centering
\begin{minipage}[b]{0.48\textwidth}
\centering
\small
\begingroup
\newcommand{\sweepermodelicon}[1]{\makebox[1.35em][l]{\raisebox{-0.18ex}{\includegraphics[width=1em,height=1em,keepaspectratio]{icons/#1.pdf}}}}
\newcommand{\sweepermodelname}[2]{\sweepermodelicon{#1}#2}
\setlength{\tabcolsep}{4pt}
\renewcommand{\arraystretch}{1.15}
\global\setbox\harnesstabbox=\hbox{\begin{tabular}{llc}
\toprule
Model & Harness & Pass \\
\midrule
\multirow{5}{*}{\sweepermodelname{openai}{GPT-5.6 Luna}} & \sweepermodelname{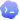}{Codex} & \textbf{42.5} \\
& \sweepermodelname{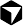}{Cursor} & 40.0 \\
& \sweepermodelname{kimi}{Kimi Code} & 37.5 \\
& \sweepermodelname{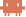}{Claude Code} & 35.0 \\
& \sweepermodelname{meta}{Muse Code} & 35.0 \\
\bottomrule
\end{tabular}}\usebox{\harnesstabbox}
\captionof{table}{\textbf{GPT-5.6 Luna does best in its native harness, Codex.} Pass rate (\%) by harness.}
\label{tab:harness}
\endgroup
\end{minipage}\hfill
\begin{minipage}[b]{0.48\textwidth}
\centering
\includegraphics[height=\dimexpr\ht\harnesstabbox+\dp\harnesstabbox\relax]{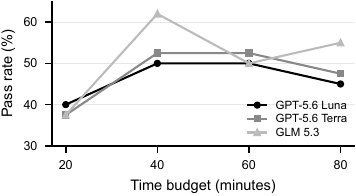}
\caption{\textbf{More time helps only up to a point.} Pass rate (\%) by time budget on 40 tasks.}
\label{fig:time-budget}
\end{minipage}
\end{figure}

\secrefc{sec:behavior} compares agents that work for different lengths of time. To test whether more time helps the same agent, we give GPT-5.6 Luna, GPT-5.6 Terra, and GLM 5.3 budgets of 20, 40, 60, and 80 minutes on the same 40 tasks. Each agent is told to keep working until its deadline. Pass rates rise from 20 to 40 minutes but not beyond (\figref{fig:time-budget}). They peak at 50.0\%, 52.5\%, and 62.0\%, respectively. With 40 tasks these differences are not statistically significant, but they give no sign that more time alone closes the gap.

\section{Related Work}

\paragraph{Software engineering agents and benchmarks.} Repository-level agents improve through better interfaces, code localization, and structured repair workflows~\citep{zhang2024autocoderover,yang2024sweagent,xia2024agentless,wang2025openhands,pan2024training,applis2026unified,lian2026swe,tian2026agent,ganhotra2026resolving,he2026swe,kuang2026reagent}. SWE-bench~\citep{jimenez2024swe} evaluates them on real GitHub issues, and its successors raise difficulty~\citep{deng2025swe,huang2026deepswe,seniorswebench2026,zheng2026swe}, add visual issues and more languages~\citep{yang2024swe,rashid2025swe,zan2026multi,shi2026swe}, refresh tasks~\citep{zhang2026swe}, and cover freelance work, broader engineering tasks, user feedback, and rebuilding programs~\citep{miserendino2025swe,raghavendra2026sweatlasbenchmarkingcoding,raghavendra2026swe,yang2026programbench}. Web, desktop, and command-line benchmarks likewise specify the goal~\citep{zhou2023webarena,OSWorld,merrill2026terminal}. In all of these, the task states what to fix or achieve, so they evaluate the stages after a bug report. \bench{} names only a product area and evaluates the step that produces the report.

\paragraph{Testing and verification.} SWT-Bench~\citep{mundler2024swtbench} asks agents to write tests that reproduce a reported issue, so the bug is known in advance. WebProber~\citep{ye2025webprober} lets an agent explore websites like a user and report usability issues, but does not evaluate repairs. In \bench{}, agents must design tests before any bug is known and then fix what they find. To grade repairs, most benchmarks run hidden unit tests written with the reference fix~\citep{jimenez2024swe,rashid2025swe,zan2026multi}, and these tests can reject valid alternative fixes. Others use engineer-written browser tests~\citep{miserendino2025swe} or scripts in the task environment~\citep{huang2026deepswe,merrill2026terminal}. Browser agents~\citep{browser_use2024} make it possible to check behavior through the interface; our agentic verifier follows step-by-step behavior tests and accepts any fix that restores what a user would observe.

\paragraph{AI code review.} Finding bugs before users do is increasingly delegated to code review agents. Early work learned to review code from data~\citep{li2022automating,frommgen2024resolving}. Tools such as Cursor Bugbot~\citep{cursor2025bugbot}, GitHub Copilot code review~\citep{github2025copilotreview}, CodeRabbit~\citep{coderabbit2026}, Greptile~\citep{greptile2026}, Codex code review~\citep{openai2026codexreview}, Claude Code Review~\citep{anthropic2026codereview}, and Gemini Code Assist~\citep{google2026geminicodeassist} now review pull requests for bugs as they are opened, and benchmarks score their comments on real pull requests~\citep{zeng2025swrbench,zhang2026sphinx}. Like \bench{}, these tools look for problems no one has reported, but they mostly analyze the diff and surrounding code. \bench{} evaluates the harder case where bugs appear only when the running application goes through a sequence of user actions.

\section{Conclusion and Discussion}

Current agents fix bugs reliably once given a bug report, yet pass at most 59.0\% of \bench{} tasks when they must find the bug themselves. The gap lies in testing, not repair. Agents must choose which scenarios to try, run them through the interface, and recognize a failure when they see one. Recognition is often where they fall short. On the Focalboard task, all four agents we inspect see the failure, but only two treat it as a bug.

This testing gap is likely to matter more as agents build software on the fly for individual users. Such software rarely passes through a testing team, so if the agent does not find a bug, the user will. Our results suggest that closing the gap requires agents that use the application the way a user would and judge its behavior by what it is meant to do, not only by what its tests expect. Verifiers that check user-observable behavior, like ours, could also reward this kind of testing during training. As agents take on more of building software, it is worth asking more of them than fixing the bugs we report: can they find those bugs before we do?

\bench{} has limitations. It covers only web applications and scores one known bug per task, so it measures recall rather than precision. Fixes to other real issues earn no credit, and unnecessary changes go unpenalized unless they break a preservation test. Its behavior tests check functional correctness only, not visual design, ease of use, or responsiveness. Extending the benchmark to other interactive software and to these aspects of user experience is a natural next step.

\subsection*{Ethics statement}

    \bench{} is constructed from publicly available open-source software projects in accordance with their licenses. All agent experiments run in isolated environments and do not affect the original repositories, deployed products, or real user data. The benchmark focuses on bugs triggered through normal user interactions rather than the discovery or exploitation of security vulnerabilities, and is not intended to facilitate offensive use. This work does not involve human subjects or private user data.

\subsection*{Acknowledgments}
We thank Modal for providing the cloud sandboxes in which we ran all agent and verifier experiments.

\bibliographystyle{plainnat}
\bibliography{references}

\clearpage
\appendix
\section*{\Large Appendix}

\section{Prompts}
\label{app:prompts}

Every agent receives the prompt below, with \texttt{\{\{TASK\_DESCRIPTION\}\}} replaced by the task's open-ended request. In the ablation of \secrefc{sec:browser-prompt}, we remove the bullet that asks the agent to test in the browser, the bullet that mentions Chromium and Playwright, and the verification bullet about opening the product UI. For the detailed prompt of \secrefc{sec:detailed-prompt}, the task description is replaced by a description of the target bug; the rest of the prompt is unchanged.

\begin{lstlisting}
You are working in `/workspace/product`.

Please perform the following task:

{{TASK_DESCRIPTION}}
Important notes:
- Use only this checkout and these instructions. Do not search external repositories or known fixes.
- Do not fetch Git history, add remotes, or commit.
- This environment has no outbound internet for packages or source. npm, PyPI, GitHub, CDNs, and similar hosts are blocked and will return 403. Do not install, fetch, download, or rebuild anything from the network. Dependencies and tools are already on disk. If a command fails with a network or 403 error, treat that as final: do not retry, switch mirrors, or keep waiting. Continue with the local checkout and already-installed tools only.
- The product server is published on this machine at 127.0.0.1:13200.
- Use the browser to do end-to-end testing to discover and test issues. Drive the product UI like a real user in the tasked area and look for bugs that show up during normal use. Do not stop after the first issue you can patch; keep exploring other screens and controls in the same area.
- Chromium and Playwright are preinstalled.

How to verify your work:
- Product commands must run in the product container through `/workspace/task-shell`.
- Open the product UI at http://127.0.0.1:13200 with the preinstalled Playwright/Chromium in this environment. Do not launch the browser through `/workspace/task-shell`.
- Do not call `yarn`, `pnpm`, or `npm` directly; use the `product-*` wrappers in the container PATH.
- Find the right test or build scripts from package.json, README, or existing CI, then run them through `/workspace/task-shell`.

Deliverable:
- Modify product source in place and leave a clean, reviewable diff with at least one real code change.
- An empty diff is not acceptable. Do not commit.
\end{lstlisting}

For the Focalboard task, the two task descriptions are:

\begin{lstlisting}
Open-ended: Find and fix issues in Focalboard's board sharing and membership.

Detailed: Fix Focalboard board sharing so adding a member via Share also puts that board in the member's default sidebar category. After a user shares a board with another member, that member should see the board under Boards in their left sidebar.
\end{lstlisting}

\section{Example Behavior Test}
\label{app:workflow}

\tabref{tab:focalboard-056-workflow} lists the complete target behavior test for the Focalboard task in \figref{fig:focalboard-procedure}, reproduced verbatim from the benchmark. The verifier follows these steps in the browser and returns a verdict from the final step. Setup problems, such as a failed login, are reported separately from product failures.

\begingroup
\small
\setlength{\tabcolsep}{5pt}
\renewcommand{\arraystretch}{1.12}
\begin{longtable}{@{}p{0.065\textwidth}p{0.905\textwidth}@{}}
\toprule
Step & Evaluation instruction \\
\midrule
\endfirsthead
\multicolumn{2}{l}{Table~\ref{tab:focalboard-056-workflow} (continued)}\\
\toprule
Step & Evaluation instruction \\
\midrule
\endhead
\midrule
\multicolumn{2}{r}{Continued on next page}\\
\endfoot
\bottomrule
\caption{\textbf{Target behavior test for the Focalboard task.} The verifier follows these instructions in the browser. Wording is reproduced verbatim; only typesetting and line wrapping are changed.}\label{tab:focalboard-056-workflow}\\
\endlastfoot
\multicolumn{2}{@{}p{0.99\textwidth}@{}}{\textbf{Target behavior test}: After miraDunn shares a newly created board with kadeRue via Share, kadeRue's left sidebar Boards category must list that board. On the buggy build kadeRue is a member but the board never appears in his sidebar.}\\
\midrule
1 & Open exactly http://127.0.0.1:13200/login and wait until the official Focalboard Log in page is visible: a styled rounded login card (not a bare unstyled text shell), the heading Log in, the username field with placeholder Enter username (id login-username), the password field with placeholder Enter password (id login-password), a filled Log in submit button, and the link or create an account if you don't have one. \\
\midrule
2 & Type exactly miraDunn into the username field. Type exactly Slate6quay into the password field. Click the Log in button. Wait until the signed-in Focalboard chrome loads: the left sidebar shows the Focalboard logo wordmark and version, and + Add board. If a Welcome To Boards page appears, click No thanks, I'll figure it out myself. If login fails or the sidebar never appears, stop and treat that as a setup problem, not a product pass/fail. \\
\midrule
3 & In the left sidebar click + Add board. Wait until the official Create a board template selector is visible (heading Create a board, templates on the left, preview on the right, buttons Use this template and Create empty board). Click Create empty board. Wait until a new board canvas opens with the title placeholder Untitled board and a top-right Share button. \\
\midrule
4 & Click the Untitled board title (the large board title, not the view name Board view). Type exactly Q3 Launch Desk and press Enter or click away so the title saves. Confirm the board title now reads Q3 Launch Desk and that the same name appears under the sidebar Boards category. If the empty board never opened or the title would not save, stop and treat that as a setup problem. \\
\midrule
5 & Click the top-right Share button (title Share board). Wait until the Share Board dialog is visible with the search placeholder Search for people and channels. Click that search field, type exactly kadeRue, and wait until a people result for @kadeRue appears. Click the @kadeRue result so kadeRue is added. Wait until the member list in the same dialog shows kadeRue (not only miraDunn / You). Close the dialog. If kadeRue cannot be found or added, stop and treat that as a setup problem, not the sidebar bug. \\
\midrule
6 & Open the Focalboard logo menu at the top of the left sidebar (the wordmark Focalboard). Click Log out. Wait until the Log in page is visible again. \\
\midrule
7 & Type exactly kadeRue into the username field. Type exactly Moss3vale into the password field. Click Log in. Wait until kadeRue's signed-in Focalboard chrome loads (sidebar Focalboard logo and + Add board). If Welcome To Boards appears, click No thanks, I'll figure it out myself. Use the browser reload/refresh once on this signed-in page so the sidebar refetches categories. If login fails, stop as a setup problem. \\
\midrule
8 & After that refresh, look only at kadeRue's left sidebar list under the Boards category (the system category named Boards). Do not open Find Boards / the board switcher (Ctrl/Cmd+K) and do not paste a board URL. Judge only whether the sidebar category lists a board titled Q3 Launch Desk. \\
\midrule
9 & Final verdict --- FAIL only if kadeRue is signed in, the left sidebar Boards category is visible, and Q3 Launch Desk is absent from that sidebar list (the shared board never landed in kadeRue's default category). PASS only if that same sidebar Boards category lists Q3 Launch Desk. Do not PASS because miraDunn could share, or because kadeRue can log in. Do not FAIL for login/navigation/template-selector problems. Do not use fetch, XHR, DevTools Network, API JSON, or database reads for setup or pass/fail. \\
\addlinespace
\end{longtable}
\endgroup

\end{document}